\documentstyle[12pt]{article}
\begin{document}
\begin{center}
{\Large{\bf Gauge Invariances in Second Class Constrained 
Systems - a comparitive look at two methods}}

\vspace{12mm}
{\large{\em A S Vytheeswaran}}{\footnote{\bf email : 
{}~ctsguest@cts.iisc.ernet.in ; ~~buniphy@kar.nic.in}}

\vspace{5mm}
{{\sf Department of Physics, \\
Bangalore University, Jnana Bharathi Campus,\\
Bangalore 560 056, ~INDIA}}
\end{center}
\begin{abstract}
We look at and compare two different methods developed earlier for 
inducing gauge invariances in systems with second class constraints.
These two methods, the Batalin-Fradkin method and the Gauge Unfixing 
method, are applied to a number of systems. We find that the extra 
field introduced in the Batalin-Fradkin method can actually be found 
within the original phase space itself.
\end{abstract}

\vspace{5mm}
\section{Introduction}

In recent years there have been a number of papers on gauge invariances 
in systems with second class constraints. Basically this involves 
unravelling, using the language of constaints, gauge symmetries hidden 
in such systems. By doing so it has sometimes been possible to obtain  
a deeper and more illuminating view of these systems. 

In unravelling such hidden symmetries, the basic idea is that the
original system is a gauge fixed version of a certain gauge theory; 
the latter reverts to the former under certain gauge fixing 
conditions. The advantage in having a gauge theory lies in the fact
that other gauges can also be considered. Further it is possible to get 
more than one gauge theory for the same second class constrained system.
 
Two methods have been suggested to make this conversion of second class 
theories into gauge theories. One method, based originally on the idea 
of Faddeev and Shatashvili \cite{FaSha}, is now called the Batalin - 
Fradkin method \cite{BaFa} and is formulated in an enlarged phase space. 
The other method, based on the idea of Mitra-Rajaraman \cite{RaPa} is 
what we call {\em gauge unfixing} \cite{RaVy}, and does not use any 
enlarged phase space. Rather it confines itself to the phase space of 
the original second class system. 

Even though these methods look very different in their formulations,
when they are applied to many systems like Chern - Simons theory, 
chiral Schwinger model, etc., the results are basically the same, 
implying that as far as these examples are concerned, the two methods 
are essentially equivalent. In what follows we illustrate this equivalence 
for three such examples. We compare the results of the two methods for 
these systems. Any conclusions arising out of such a comparison might be 
illuminating when the formal equivalence is considered. Such a formal 
equivalence will be considered separately.

In Section 2 we review the two methods, and look at specific 
systems in Section 3. We conclude in Section 4.

\section{The Formalisms}

We consider a finite dimensional system in phase space with 
co-ordinates $ q^i $ and conjugate momenta $ p_i ~(i = 1,2,
\ldots N). $ The system has two second class constraints,
\begin{equation}
Q_1(q,p) \approx 0, \hspace{1.2in} Q_2(q,p) \approx 0
\end{equation}
defining a constraint surface $ \sum_2$. Due to their second 
class nature, the $2 \times 2$ antisymmetric matrix $E$ whose 
elements $E_{mn}$ are Poisson brackets among the $ Q$'s,
\begin{equation}
E_{ab}(q,p) = \{ Q_a, Q_b \}\hspace{0.8in}a,b=1,2
\end{equation}
is {\em invertible} everywhere, even on the surface $ \sum_2. $ 
The canonical Hamiltonian is $ H_c $ and the total Hamiltonian is 
\begin{equation}
H = H_c + \mu_1Q_1 + \mu_2Q_2,
\end{equation} 
where the multipliers $ \mu_1, \mu_2  $ are determined on the 
surface $ \sum_2 $ by demanding consistency of the two constraints 
with respect to $H$. 

\subsection{Batalin - Fradkin (BF) method}

As mentioned earlier, the idea behind this method \cite{FaSha, BaFa} 
is to enlarge the phase space by including new variables. Since we have 
taken the number of second class constraints here to be two, we 
introduce two variables $\Phi^a (a=1,2)$. The enlarged phase space 
$(q,p,\Phi)$ has the basic Poisson brackets
\begin{equation}
\{ q^i, p_j \} = \delta^i_j \hspace{1.2in} \{\Phi^a, \Phi^b\} = 
\omega^{ab}
\end{equation}
with all other Poisson brackets zero. The antisymmetric $ 2\times 
2 $ matrix $ \omega^{ab} $ is a constant matrix, unspecified for 
the present.

The first class constraints are obtained as functions in this {\em 
extended} phase space. Since we had initially two second class 
constraints, there will now be {\em two} first class constraints, 
given in general by
\begin{eqnarray}
{\widetilde Q_a}(q^i,p_i,\Phi^a) & = & Q_a + \sum_{m = 1}^{
\infty} Q_a^{(m)}, \hspace{0.8in} Q_a^{(m)} \sim (\Phi^a)^{m} \\
{\widetilde Q_a}(q^i,p_i,0) & = & Q_a^{(0)} = Q_a \nonumber
\end{eqnarray}
where the second line gives the boundary condition. The terms of 
various orders in the expansion for $ {\widetilde Q_a} $ are 
obtained by demanding that the $ {\widetilde Q_a} $ are strongly 
first class,
\begin{equation}
\{ {\widetilde Q_a}, {\widetilde Q_b} \} = 0.
\end{equation}
For instance for the first order this requirement gives  
\begin{equation}
E_{ab} + X_{ac}\omega^{cd}X_{db} = 0
\end{equation}
which can be satisfied, using (4), if we write and substitute
\begin{equation}
Q_a^{(1)} = X_{ab}(q,p)\Phi^b,
\end{equation}
in (6) and consider terms upto first order. For many systems, 
such as the ones we consider in the next section, the higher 
order terms are all zero. It must be noted that there is an 
inherent {\em arbitrariness} in the choice of the $ \Phi^a $ 
and the coefficients $ X_{ab}. $ This choice may be exploited 
to advantage.

To get gauge invariant observables, we note that relevant 
quantities of the original second class system in general are 
not gauge invariant with respect to the new first class constraints. 
They are made gauge invariant by modifying them in the extended 
phase space. In particular the gauge invariant Hamiltonian \cite{BaFa} 
can be written in general as 
\begin{equation}
{\widetilde H} = H + \sum_{m=1}^{\infty} H^{(m)} \hspace{1in}
H^{(m)} \sim (\Phi^a)^{m}
\end{equation}
and the terms of various orders are obtained by demanding that
\begin{equation}
\{ {\tilde H}, {\widetilde Q_m} \} = 0.
\end{equation}
A similar procedure is followed to obtain other gauge invariant quantities
also \cite{BaFa}. We finally remark that eqn. (7) can always be written so 
in the case of 2 constraints. For more than 2 second class constraints, 
this has to be taken as an assumption, which {\em need not} hold in the 
very general case. In a sense, the $ X_{ab} $ can be called the ``square 
root'' of the matrix $E$ \cite{BaFa}.
       
\subsection{The Gauge Unfixing method}

This method \cite{RaPa,RaVy}, in contrast to the BF method, makes no 
enlargement of the phase space. Rather, since the number of second 
class constraints is even (we consider here only bosonic constraints), 
only half of these constraints are chosen to form a first class subset, 
and the other half as the corresponding gauge fixing subset. This latter 
subset is discarded, retaining only the first class subset, and so we 
have a gauge theory. 

In a general system, getting a first class subset is a non-trivial 
issue \cite{RaVy}; this might be possible only under certain conditions. 
However in the case of only two second class constraints (which we 
consider here) the first class constraint can always be chosen.

For instance, we can choose $ Q_1 $ as our first class constraint, 
and $ Q_2 $ as its gauge fixing constraint. We redefine, using (2),
\begin{equation}
Q_1 \rightarrow \chi = E_{12}^{-1}Q_1 \hspace{1.5in} Q_2 
= \psi
\end{equation}
and no longer consider the $ \psi $ as a constraint. The gauge invariant 
Hamiltonian and other physical quantities are obtained by defining 
the projection operator
\begin{equation}
I\!\!P \cong ~: e^{-\psi{\hat \chi}} : \hspace{1.2in} {\hat \chi}(A) \equiv
\{ \chi, A \}
\end{equation}
and operating $I\!\!P$ on any phase space function $A$. There is 
an ordering prescribed for the action of $ I\!\!P; $ the $ \psi 
$ is always outside the Poisson bracket in the expansion of the 
exponential. The action of $ I\!\!P $ on relevant quantities gives 
the gauge invariant quantities.

It must be noted that even in this method, there is an inherent {\em 
arbitrariness}; either of the two second class constraints can be chosen 
to be first class. The two choices define two different projection 
operators, and the gauge theories so constructed will in general be 
different. This arbitrariness can be exploited to advantage.

\section{Examples} 
\subsection{Chiral Schwinger Model}

This well known anomalous gauge theory \cite{JaRa,Kim,Asv} involves 
chiral fermions coupled to a $U(1) $ gauge field in $1+1$ dimensions. 
Classically the theory has gauge invariance, but this is lost upon 
quantisation. We look at its bosonised version, the advantage being 
that the corresponding classical theory itself has no gauge invariance. 
We have
\begin{equation}
{\cal L} = -\frac{1}{4}F_{\mu\nu}F^{\mu\nu} + \frac{1}{2}(\partial_{\mu}
\phi)^2 + e(g^{\mu\nu} - \epsilon^{\mu\nu})(\partial_{\mu}\phi)A_{\nu} 
+ \frac{1}{2}e^2\alpha A_{\mu}^2
\end{equation}
where $g^{\mu\nu} $ = diag$(1,-1)$, ~$ \epsilon^{01} = - \epsilon^{10} = 
1 $ and $ \alpha $ is the regularisation parameter. The theory is gauge 
non-invariant for all values of $ \alpha. $ We consider the case $ \alpha 
> 1.$

The canonical Hamiltonian density is
\begin{eqnarray}
{\cal H}_c & = & \frac{1}{2}\pi_1^2 + \frac{1}{2}\pi_{\phi}^2 + \frac{1}{2}
(\partial_{1}\phi)^2 + e(\partial_1\phi + \pi_{\phi})A_1 + \frac{1}{2}
e^2(\alpha + 1)A^2_{1} \nonumber \\
&& \rule{0mm}{8mm}\hspace{0.6in} - A_0\left [- \partial_1\pi_1 + 
\frac{1}{2}e^2(\alpha - 1)A_0 + e(\partial_1\phi + \pi_{\phi}) + e^2A_1 
\right ]
\end{eqnarray}
where $ ~\pi_1 = F^{01} = \partial^0A^1 - \partial^1A^0, $ and $ \pi_{\phi} 
= \partial_0\phi + e(A_0 - A_1) $ are the momenta conjugate to $ A_1 $ and 
$\phi$ respectively. The constraints are 
\begin{eqnarray}
Q_1 & = & \pi_0 = 0\nonumber \\
Q_2 & = & - \partial_1\pi_1 + e^2(\alpha - 1)A_0 + e(\partial_1\phi + 
\pi_{\phi}) + e^2A_1 = 0
\end{eqnarray}
defining a constraint surface $ \sum_2. $ These are of the second class, 
\begin{equation}
E_{12} = \{ Q_1(x), Q_2(y) \} = -e^2(\alpha - 1)\delta(x - y).
\end{equation}

\vspace{2mm}
Following the BF method \cite{Kim}, the phase space is extended by 
introducing two fields $ \Phi_1, \Phi_2 $, with Poisson bracket relations 
of the form (4). To get the first class constraints (5) and (8), we recall 
that there is a natural arbitrariness in choosing the matrices $ 
\omega^{ab} $ and $ X_{ab}. $ This allows the choice  
\begin{eqnarray}
\omega = \left( \begin{array}{c}
        0 ~~~ 1\\
       -1 ~~ 0 
\end{array}                   \right) \delta(x-y) \hspace{0.6in}
X (x,y) = e\sqrt{\alpha-1}\left( \begin{array}{c}
                                1 ~~ 0\\
                                0 ~~ 1 
\end{array}             \right) \delta(x-y)
\end{eqnarray}

\vspace{1mm}
\noindent This choice allows the two new fields to form a canonically 
conjugate pair. The terms beyond the first order in the expansion (5) 
are all zero. We then get
\begin{equation}
{\widetilde Q_m} = Q_m + e\sqrt{\alpha-1}\Phi_m, \hspace{0.8in}m = 1,2
\end{equation}
which, using (16) and (17), can be verified to be strongly first class.

Using similar arguments the gauge invariant Hamiltonian for the choice 
(17) is 
\begin{eqnarray}
{\widetilde H}_{BF} 
& = & H_c + {\displaystyle \int}dx\left[ - \frac{\displaystyle e\pi_1 + 
e(\alpha-1)\partial_1A_1}{\displaystyle \sqrt{\alpha-1}} \Phi_1 + 
\frac{\displaystyle e^2}{\displaystyle 
2(\alpha-1)}\Phi_1^2 \right.\nonumber \\
&& \hspace{1.6in}\left.+ \frac{1}{2}(\partial_1\Phi_1)^2  + \frac{1}{2}(
\Phi_2)^2 - \frac{{\widetilde Q_2}\Phi_2}{\displaystyle e\sqrt{\alpha-1}} 
\right]
\end{eqnarray}
where $ H_c $ is the canonical Hamiltonian. This $ {\widetilde H_{BF}} 
$ has zero PBs with the first class constraints (18).

\vspace{4mm}
Coming to the {\em Gauge Unfixing} (GU) method \cite{Asv}, we reiterate 
that no new field need be introduced. The first class constraint is taken 
to be just one of the two existing constraints. We choose, after a 
rescaling
\begin{equation}
\chi = \frac{1}{\displaystyle e^2(\alpha-1)} Q_2 ~,
\end{equation}
so that the relevant constraint surface is $ \sum_1 $ defined by $ \chi 
\cong 0.$ The gauge fixing-like constraint is $ \psi = 0,$ and is discarded
(that is {\em unfixed}). To get the gauge invariant Hamiltonian we 
construct a projection operator $ I\!\!P $ of the form (12) and use it on 
the canonical Hamiltonian $H_c$,
\begin{equation}
{\widetilde H}_{GU} = H_c + {\displaystyle \int} 
dx\left[\frac{\displaystyle \pi_1 + (\alpha-1)\partial_1A_1}{\displaystyle 
{\alpha-1}} Q_1 + \frac{1}{\displaystyle 2e^2(\alpha-1)}(\partial_1Q_1)^2 
+ \frac{1}{\displaystyle 2(\alpha-1)^2}Q_1^2, \right]
\end{equation}
which is gauge invariant with respect to the $\chi$.

We see that, apart from the term $ {\displaystyle \int} dx \left 
(\frac{\displaystyle \Phi_2^2}{2} - \frac{\displaystyle \Phi_2}{
\displaystyle 2\sqrt{\alpha - 1}}{\widetilde Q_2} \right), $ the $ 
{\widetilde H_{GF}}$ and the ${\widetilde H_{GU}}$ in (19) and (21) are 
the same, if we make the identification $ \Phi_1 = - \frac{\displaystyle 
Q_1}{\displaystyle e\sqrt{\alpha - 1}} $. We however emphasise the two 
rather different paths used to get these Hamiltonians. One requires the 
introduction of an extra (canonical) pair of fields, while the other 
doesn't need this. In both cases extra terms are needed to make the 
original Hamiltonian gauge invariant. For the $ {\widetilde H_{BF}} $ 
these terms had to be written down by introducing extra fields, whereas 
in the $ {\widetilde H_{GU}} $ these terms involve a variable 
{\em already present} in the original theory.

\vspace{4mm}
We look at the path integral quantisation for these two Hamiltonians. For 
Hamiltonian $ {\widetilde H_{BF}} $ we have,
\begin{eqnarray}
{\cal Z}_{BF} & = & {\displaystyle \int}{\cal D}(\pi_{\mu}, A^{\mu},
\pi_{\phi}, \phi, \Phi_1, \Phi_2, \mu_1, \mu_2) ~e^{iS_{BF}}\\
S_{BF} & = &  {\displaystyle 
\int} dxdt \left [\pi_0{\dot A^0} + \pi_1{\dot A^1} + \pi_{\phi}{\dot 
\phi} + \Phi_2{\dot \Phi_1} - {\widetilde{\cal H}}_{BF} - \mu_1{\widetilde 
Q_1} - \mu_2{\widetilde Q_2} \right].\nonumber
\end{eqnarray}
Here $ \mu_1, \mu_2 $ are undetermined Lagrange multipliers 
corresponding to the first class constraints $ {\widetilde Q_1}, {
\widetilde Q_2} $ respectively. If we make the transformations $ \mu_2 
\rightarrow \mu_2^{\prime} = \mu_2 - \frac{\displaystyle \Phi_2}{
\displaystyle e(\alpha - 1)}, $
$ A_0 \rightarrow A_0^{\prime} = A_0 - \mu_2^{\prime}, ~\pi_1 
\rightarrow \pi_1^{\prime} = \pi_1 + \partial_0A_1 - \partial_1A_0^{
\prime}, ~\pi_{\phi} \rightarrow \pi_{\phi}^{\prime} = \pi_{\phi} - {\dot 
\phi} - e(A_0^{\prime} - A_1), $ and then integrate over $ \pi_i^{\prime}, 
{}~\pi_{\phi}^{\prime}, ~\Phi_2, $ we get
\begin{eqnarray}
{\cal Z}_{BF} & = & {\displaystyle \int} {\cal D}(A^{\mu}, \phi, \Phi_1) 
{}~~e^{iS_{BF}} \nonumber \\
S_{BF} & = & {\displaystyle \int} dxdt \left( - \frac{1}{4}F_{\mu\nu}
F^{\mu\nu} + \frac{e^2\alpha}{2}A_{\mu}A^{\mu} + e(\eta^{\mu\nu} - 
\epsilon^{\mu\nu})(\partial_{\mu}\phi)A_{\nu} + 
\frac{1}{2}(\partial_{\mu}\phi)^2  \right.\\
&& \hspace{0.8in} \left. + \frac{1}{2}(\partial_{\mu}\Phi_1)^2 - 
\frac{e}{\sqrt{\alpha-1}}\Phi_1[(\alpha-1)\eta^{\mu\nu} 
+ \epsilon^{\mu\nu}](\partial_{\mu}A_{\nu})\right).
\nonumber
\end{eqnarray}
The action $ S_{BF} $ above is just the gauge invariant version for the 
chiral Schwinger model. As is well known, this action was first obtained 
by merely adding the (Schwinger) terms in the variable $ \Phi_1 $ to 
the original bosonised action (13). It has also been obtained using 
other arguments. In the Batalin-Fradkin approach, these Schwinger terms 
and $ \Phi_1 $ come up due to the extension of the phase space.

\vspace{4mm}
Coming to the Hamiltonian $ {\widetilde H_{GU}} $ of the Gauge Unfixing
method, we have
\begin{eqnarray}
{\cal Z}_{GU} & = & {\displaystyle \int} {\cal D}(A^{\mu}, \pi_{\mu}, 
\phi, \pi_{\phi}, \lambda) ~e^{iS_{GU}}\nonumber\\
S_{GU} & = & {\displaystyle \int} dxdt \left [\pi_0{\dot A^0} + 
\pi_1{\dot A^1} + \pi_{\phi}{\dot \phi} - {\cal {\widetilde H}}_{GU} - 
\lambda\chi \right],
\end{eqnarray}
where $ \lambda $ is the arbtirary Lagrange multiplier. We make the 
transformations $ A_0 \rightarrow A_0^{\prime} = A_0 - \frac{\lambda}{
e^2(\alpha-1)}, ~\pi_1 \rightarrow \pi_1^{\prime} = \pi_1 + \partial_0
A_1 - \partial_1A_0^{\prime} + \frac{\pi_0}{\alpha-1}, \pi_{\phi}
\rightarrow \pi_{\phi}^{\prime} = \pi_{\phi} - {\dot \phi} + eA_1 - 
eA_0^{\prime}$ and $ \lambda \rightarrow \lambda^{\prime} = \lambda 
+ \partial_0\pi_0. $ We then integrate over $ \pi_1^{\prime}, \pi_{\phi
}^{\prime}$ and $ \lambda^{\prime} $ in the path integral to get
\begin{eqnarray}
{\cal Z}_{GU} & = & {\displaystyle \int} {\cal D}(A^{\mu}, \phi, \pi_0) 
{}~e^{iS_{GU}}\nonumber\\
S_{GU} & = & {\displaystyle \int} dxdt \left( - \frac{1}{4}F_{\mu\nu}
F^{\mu\nu} + \frac{e^2\alpha}{2}A_{\mu}A^{\mu} + e(\eta^{\mu\nu} - 
\epsilon^{\mu\nu})(\partial_{\mu}\phi)A_{\nu} + \frac{1}{2}(\partial_{
\mu}\phi)^2  \right.\\
&& \hspace{1.2in} \left. + \frac{1}{2}\frac{\displaystyle (\partial_{\mu}
\pi_0)^2}{\displaystyle e^2(\alpha-1)} + \frac{\displaystyle  
\pi_0}{\displaystyle \alpha-1}[(\alpha-1)\eta^{\mu\nu} + 
\epsilon^{\mu\nu}](\partial_{\mu}A_{\nu})\right).\nonumber
\end{eqnarray}
We see that on making the replacement $ \pi_0 = - e\sqrt{\alpha-1}\Phi_1  
$ in (25), we get the same result as in the Batalin-Fradkin case (23).
This is achieved here by introducing no extra fields. The extra field 
of the BF method is found here {\em within} the original phase space. 
Further the Schwinger terms are the same in both cases. We also note  
that the extra term $ \int \left(\frac{(\Phi_2)^2}{2} + \ldots \right)$ 
{} which comes upon comparing the Hamiltonians in (19) and (21) have been
integrated away in the path integral (23).

\subsection{The abelian Proca model}

This $ (3+1) $ - dimensional theory is given by the Lagrangian \cite{Stu,
NbRb, AsvP}
\begin{equation}
{\cal L} = -\frac{1}{4}F_{\mu\nu}F^{\mu\nu} + \frac{m^2}{2}A^{\mu}A_{\mu}
\end{equation}
where $m$ is the mass of the $ A_{\mu} $ field, $ g_{\mu\nu} = $diag$ ~
(+,-,-,-)$
and $ F_{\mu\nu} = \partial_{\mu}A_{\nu} - \partial_{\nu}A_{\mu}. $ The 
canonical Hamiltonian is given by 
\begin{equation}
H_c = {\displaystyle \int d^3x ~{\cal H}_c} =  {\displaystyle \int 
d^3x \left ( \frac{1}{2}{\pi_i}{\pi_i} 
+ {\frac{1}{4}}F_{ij}F_{ij} - \frac{m^2}{2}(A_0^2 - A_i^2) + 
A_0({\partial_i}\pi_i)\right ),}
\end{equation}
with $ \pi_i = -F_{0\,i} $ the momenta conjugate to the $A^i$. The second 
class constraints are
\begin{equation}
Q_1 = \pi_0(x) \approx 0,\hspace{1.2in}
Q_2 = (-{\partial_i}\pi_i + {m^2}A_0)(x) \approx 0,
\end{equation}
which together define the surface $ \sum_2 $ in the phase space. Their 
second class nature is due to the mass term in the Lagrangian. 
The matrix $E$ of eqn. (2) is here
\begin{equation}
E =      \left( \begin{array}{c}
             0\;\;\;{-m^2}\\
               \!\!{m^2}\:\;\;\;0
                \end{array}\right ) \delta(x-y),
\end{equation}
whose determinant is non-zero everywhere.

\vspace{3mm}
Using the Batalin - Fradkin method \cite{NbRb}, we introduce an extra 
canonical pair of fields $ \theta $ and $ \pi_{\theta}, $ with $ \{
\theta(x), \pi_{\theta}(y) \} = \delta(x-y). $ As earlier, transformations 
in this extended phase space modify the constraints (28) to 
\begin{equation}
{\widetilde Q_1} = Q_1 + m^2 \theta \hspace{1.2in} {\widetilde Q_2} = 
Q_2 + \pi_{\theta},
\end{equation}
which, using (29) can be seen to be strongly first class. The form for 
these constraints corresponds to the choice of $ \theta $ and $ 
\pi_{\theta} $ as a canonically conjugate pair.

The corresponding gauge invariant Hamiltonian is
\begin{equation}
{\widetilde H_{BF}} = H_c + {\displaystyle \int} d^3x \left(\frac{
\displaystyle \pi_{\theta}^2}{\displaystyle 2m^2} + \frac{m^2}{2}
(\partial_i\theta)^2 - m^2\theta\partial_iA_i\right),
\end{equation}
with respect to which the first class constraints
satisfy 
\begin{equation}
\{ {\widetilde Q_1}, {\widetilde H_{BF}} \} = {\widetilde Q_2}\hspace{1in}
\{ {\widetilde Q_2}, {\widetilde H_{BF}} \} = 0
\end{equation}

\vspace{4mm}
On the other hand, in the Gauge Unfixing method, there is only one first 
class constraint, one of the two in (28). For our purposes we choose this 
constraint to be
\begin{equation}
\chi = \frac{1}{m^2}(-\partial_i\pi_i + m^2A_0) \approx 0,
\end{equation}
and throw away the other $ \psi = \pi_0. $ The relevant constraint surface
is defined by $ \chi \approx 0. $ The $ H_c $ of (27) does not have zero 
PB with this $ \chi $ on this new surface, and hence is not gauge 
invariant. Using a projection operator of the form (12) on $H_c$ we get 
the gauge invariant Hamiltonian
\begin{equation}
{\widetilde H_{GU}} = H_c + \int d^3x \left[\psi \partial_iA_i -
\frac{1}{2m^2} \psi\partial^2_i\psi\right].
\end{equation}
Note the similarity between the Hamiltonians $ {\widetilde H_{BF}} $ and
$ {\widetilde H_{GU}}. $ Indeed, apart from the term $ \int d^3x~ \frac{
\displaystyle \pi_{\theta}^2}{2m^2} $ in (31) the two Hamiltonians are 
the same if we make the identification $ \psi = -m^2\theta. $

\vspace{4mm}
We look at path integral quantisations. For the $ {\widetilde H_{BF}} $, 
we have
\begin{eqnarray}
{\cal Z}_{BF} & = & {\displaystyle \int} {\cal D}(\pi_o, A_0, \pi_i, 
A^i, \pi_{\theta}, \theta, \mu_1, \mu_2) ~e^{iS_{BF}} \\
{S_{BF}} & = & {\displaystyle \int} d^4x \left[\pi_0{\dot A}_0 + 
\pi_i{\dot A}^i + \pi_{\theta}{\dot \theta} - {\cal H}_c - \frac{
\displaystyle \pi_{\theta}^2}{2m^2} - \frac{m^2}{2}
(\partial_i\theta)^2 + m^2\theta\partial_iA_i - \mu_1{\widetilde Q_1}
- \mu_2{\widetilde Q_2}\right],\nonumber
\end{eqnarray}
where $ \mu_1 $ and $ \mu_2 $ are undetermined Lagrange multipliers.   
After some redefinitions of fields and integration over momenta and the 
$ \mu$'s, we get
\begin{equation}
{\cal Z}_{BF} = {\displaystyle \int} {\cal D}(A^{\mu}, \theta) ~
exp ~{i{\displaystyle \int} d^4x \left[ - \frac{1}{4}
F^2_{\mu\nu} + \frac{m^2}{2} A_{\mu}^2 + \frac{m^2}{2}\partial_{\mu}
\theta\partial^{\mu}\theta - m^2\theta\partial_{\mu}A^{\mu} \right ].}
\end{equation}
The last line gives just the St\"uckelberg gauge invariant action 
\cite{Stu}. The $ \theta $ field is called the St\"uckelberg scalar, 
which was originally introduced by St\"uckelberg directly into the 
Proca Lagrangian to make it gauge invariant. Thus in the BF formalism, 
the extra field introduced is the St\"uckelberg scalar.

\vspace{4mm}
On the other hand, the path integral for the gauge unfixed Hamiltonian
$ {\widetilde H_{GU}} $ is 
\begin{equation}
{\cal Z}_{GU}  =  {\displaystyle \int} {\cal D}(A^{\mu}, \pi_{\mu}, 
\lambda)~exp ~{i\displaystyle \int} d^4x ~\left(\pi_0{\dot A_0} + 
\pi_i{\dot A_i} - {\widetilde H_{GU}} - \lambda \chi\right),
\end{equation}
with no extra fields. After redefinition of $ A_0, \pi_i $ and the $ 
\lambda $ and integrating over $ \lambda^{\prime} $ and $ \pi_i^{\prime}, 
$ we get
\begin{equation}
{\cal Z}_{GU} = {\displaystyle \int}{\cal D}(A^{\mu}, \theta) ~
exp ~i{\displaystyle \int} d^4x ~\left[ - \frac{1}{4}F_{\mu\nu}^2 
+ \frac{m^2}{2}A_{\mu}^{\mu} + \frac{m^2}{2}(\partial_{\mu}\theta)^2 - 
m^2\theta\partial_{\mu}A^{\mu} \right],
\end{equation}
where $ \theta = - \frac{\displaystyle \pi_0}{\displaystyle m^2}. $ This 
is just the path integral (37) obtained earlier using the Batalin-Fradkin 
method. 
No extra fields were introduced. Rather the extra co - ordinate field of 
the BF method which was recognised earlier as the St\"uckelberg scalar 
corresponds here to $ - \frac{\displaystyle \pi_0}{m^2}, $ which was 
already present in the phase space of the original second class 
constrained theory. This suggests that the extra field need not be 
introduced at all.

\subsection{Abelian Chern-Simons Theory}

This $ 2+1 $ dimensional theory \cite{RaVy, RaBa} consists of a 
complex field interacting with an abelian Chern-Simons field. The 
theory is described by the Lagrangian density
\begin{equation}
{\cal L} = ({\cal D}_{\mu}\phi)^*({\cal D}^{\mu}\phi) + 
\frac{\alpha}{4\pi} 
\epsilon_{\mu\nu\lambda}A^{\mu}\partial^{\nu}A^{\lambda},
\end{equation}
with $ {\cal D}_{\mu}\phi = (\partial_{\mu} - iA_{\mu})\phi, ~g_{\mu\nu}$ 
= diag $(1,-1,-1). $ and $ \mu,\nu,\lambda = 0,1,2$. We also have
\begin{equation}
H_c = \int d^2x  \left [({\vec \nabla} + i{\vec A})\phi^* \cdot ({\vec 
\nabla} - i{\vec A})\phi + \pi_{\phi^*}\pi_{\phi} + A_0(j_0 - 
\frac{\alpha}{2\pi}\epsilon_{ij}\partial_iA_j)\right ]
\end{equation}
with $ \pi_{\phi} = {\dot \phi}^* + iA_0\phi^*, ~\pi_{\phi^*} = 
{\dot \phi} - iA_0\phi,$ and $ j_0 = i(\phi\pi_{\phi} - 
\phi^*\pi_{\phi^*})$. 

\vspace{1mm}
\noindent The constraints are 
\begin{eqnarray}
\pi_0 \approx 0 &\hspace{0.1in}& Q_3 = (j_0 - \frac{\alpha}{2\pi}
\epsilon_{ij}\partial_iA_j) - \frac{\alpha}{2\pi}\partial_1Q_1 +  
\partial_2Q_2 \approx 0\nonumber\\
Q_1 = -\frac{2\pi}{\alpha}(\pi_1 + \frac{\alpha}{4\pi}A_2) \approx 0 
&\hspace{0.3in}& Q_2 = (\pi_2 - \frac{\alpha}{4\pi}A_1) \approx 0
\end{eqnarray}
with the first line showing the first class constraints. The 
second line gives the second class constraints of the theory,
\begin{equation}
\{ Q_1(x), Q_2(y) \} = \delta(x-y).
\end{equation}

Instead of the canonical Hamiltonian (40), we will consider the total
Hamiltonian which guarantees the time consistency of $Q_1$ and $Q_2$ (on 
the surface defined by both $Q_1$ and $Q_2$),
\begin{eqnarray}
H = {\displaystyle \int}d^2x \left \{{\cal H}_c + u_1Q_1 + u_2Q_2 
\right\} & \hspace{0.21in} & u_1 = i\phi^*{\cal D}_2\phi - 
i\phi({\cal D}_2\phi)^* + \frac{\alpha}{2\pi}\partial_1A_0 
\nonumber\\ 
&& u_2 = \frac{2\pi}{\alpha} \left [i\phi^*{\cal D}_1\phi - i\phi({\cal 
D}_1\phi)^* - \frac{\alpha}{2\pi}\partial_2A_0 \right ].
\end{eqnarray}

\vspace{3mm}
We now get the gauge theory using the Batalin-Fradkin method \cite{RaBa}. 
The new variables $ \Phi^1, ~\Phi^2 $ serve to enlarge the phase space, 
and have the Poisson brackets (4). In this enlarged phase space, we have 
the strongly first class constraints (after appropriate choice of the $ 
\omega $ and the $ X $ matrices),
\begin{equation}
{\widetilde Q}_1 = Q_1 + \Phi^1 \hspace{1.4in} {\widetilde Q}_2 = Q_2 
- \Phi^2.
\end{equation}
The corresponding Batalin-Fradkin gauge invariant Hamiltonian is
\begin{equation}
{\widetilde H}_{BF} = H + {\displaystyle \int} d^2x \left 
\{\phi\phi^*(\Phi^1)^2 + 
\left (\frac{2\pi}{\alpha}\right )^2 \phi\phi^*(\Phi^2)^2 - 2\phi\phi^*
\left [\Phi^1{\widetilde Q}_1 - \Phi^2{\widetilde Q}_2 \right]\right \}
\end{equation}

\vspace{2mm}
We apply the {\sf gauge unfixing} method \cite{RaVy} to this theory. We 
redefine
\begin{equation}
\chi = -\frac{2\pi}{\alpha}(\pi_1 + \frac{\alpha}{4\pi}A_2)   
\hspace{1.4in} \psi = (\pi_2 - \frac{\alpha}{4\pi}A_1),
\end{equation}
with $ \{ \chi(x), \psi(y) \}  = \delta(x-y).$ As usual we choose the $ 
\chi $ as the first class constraint and 
discard the $ \psi $. The gauge invariant Hamiltonian with respect to $ 
\chi $ is given by 
\begin{equation}
{\widetilde H}_{GU} = H + {\displaystyle \int} d^2x \left [ 
(\frac{2\pi}{\alpha})^2 \phi\phi^* \psi^2 \right ].
\end{equation}
We see that apart from the term $ \int d^2x ~\phi\phi^*(\Phi^1)^2 $ 
(and those proportional to the ${\widetilde Q}_1$ and $ {\widetilde Q}_2$ 
in (45)), the Hamiltonians $ {\widetilde H}_{BF} $ and $ {\widetilde 
H}_{GU} $ are the same. However this extra term can be easily introduced 
in the Hamiltonian $ {\widetilde H}_{GU}, $ since it is basically 
proportional to the gauge unfixing first class constraint $ \chi. $ 

By making use of these extra terms in $ {\widetilde H}_{GU}, $ we can
see this equivalence using the path integral too. After various 
redefinitions and integrations, we get the final gauge invariant action 
to be
\begin{eqnarray}
S & = & {\displaystyle \int} d^2x dt \left \{ ({\cal D}_{\mu}\phi)^*
{\cal D}^{\mu}\phi + \frac{\alpha}{4\pi}\epsilon_{\mu\nu\lambda}
A^{\mu}\partial^{\nu}A^{\lambda} - \left (\frac{2\pi}{\alpha}\right 
)^2\phi\phi^*(\Phi^2)^2 \right.\nonumber\\
&& \hspace{2.0cm}\left. - (\frac{2\pi}{\alpha}\Phi^2)\left[i\phi^*{\cal 
D}_1\phi - i\phi({\cal D}_1\phi)^* + \frac{\alpha}{2\pi} F_{02} \right] 
\right.\\
&& \hspace{1cm}\left. + \frac{1}{4\phi\phi^*}\left[{\dot \Phi^2} + 
\frac{\alpha}{2\pi}F_{01} - [i\phi^*{\cal D}_1\phi - i\phi({\cal 
D}_1\phi)^* + \frac{\alpha}{2\pi} F_{02} \right]^2\right\}.\nonumber
\end{eqnarray}
We have deliberately omitted the subscripts BF and GU here, in order to
emphasize that the same result is obtained for both the methods. The BF
action is as given above, with the terms in the $ \Phi^2 $ as extra
terms in order to give the new gauge symmetry. In the GU result the 
action is the same as in (48), with the $ \Phi^2 $ being replaced by the 
$\psi$. It may also be noted that the action in its final form is not 
manifestly Lorentz invariant.

\section{Conclusion}
In conclusion, we have seen that for the three systems above, the two
vastly different methods described in Section 2 unearth essentially the 
same gauge theories. This is seen both classically and using the path 
integral. In both methods, extra terms have to be introduced 
in the Hamiltonian; in the ${\widetilde H}_{BF} 
$ case these terms involve new fields, whereas in ${\widetilde H}_{GU}$ 
these terms are found in the original phase space. Thus, at least as far 
as the above three systems are concerned, one need not really introduce 
a totally new variable. In other words to get the hidden gauge symmetries 
one need not look outside the original system, they are present within 
the original system itself.

When their second class constraint structures are considered, the above 
three systems are simple ones. The $E$ matrices of (2) involve only 
constants, so that getting the gauge invariant Hamiltonians is quite easy; 
the new Hamiltonians will have a finite number of terms. This situation 
however need not be seen for other systems. Sometime the gauge invariant 
Hamiltonians may be in series form, in which case it has to be seen if 
closed form expressions are possible. It is to be seen if the two methods 
are equivalent in such more general cases also. This question is being 
currently looked into.

In our analysis above, the three systems involved only two second class 
constraints. It was mentioned earlier that in the gauge unfixing case, 
this does not pose any problem in choosing the first class constraint; 
either one of the two can be chosen, to give more than one gauge theory. 
However in the Batalin-Fradkin case, both constraints are to be converted 
into first class; but even here this conversion is always possible, the 
reason being the ``square root'' matrix $X$ of (8) can always be obtained 
{}from the $E$ matrix.

On the other hand, in the case of more than two second class constraints, 
we may have additional problems; in the GU method, the choice of the first 
class subset becomes non-trivial, and in the BF method finding the  
``square root'' $X$ matrix becomes non-trivial. But once the first class 
subset (or $X$ matrix) is found, then the new gauge symmetry is defined. In 
this regard we mention that in the GU method, a certain assumption regarding 
the $E$ matrix of (2) has to be used to obtain the first class subset. 

\newpage
{\noindent\large{\bf{Acknowledgements}}}

\vspace{3mm}

We wish to thank the Council for Scientific and Industrial Research,
New Delhi, India for financial assistance for this work. We also thank 
Dr B A Kagali (BU) for constant encouragement, and Prof M N Anandaram 
(BU) and Centre for Theoretical Studies (Indian Institute of Science, 
Bangalore) for providing computer facilities.


\begin{thebibliography}{999}
\bibitem{FaSha} L D Faddeev and S L Shatashvili. {\em Phys. Lett.} 
{\bf B 167} (1986) 225.
\bibitem{BaFa}  I A Batalin and E S Fradkin. {\em Phys. Lett.} {\bf B 
180} (1986) 157; {\em Nucl. Phys.} {\bf B 279} (1987), 514;
I A Batalin and I V Tyutin, {\em Int J Mod Phys} {\bf A 6} (1991), 3255.  
\bibitem{RaPa} R Rajaraman and P Mitra {\em Ann. Phys. (N Y)} {\bf 203} 
(1990), 137,157.
\bibitem{RaVy}R Anishetty and A S Vytheeswaran {\em J Phys.} {\bf A 26} 
(1993) 5613; \\
A S Vytheeswaran, {\em Ann Phys (N Y)} {\bf 236}, (1994) 297.
\bibitem{JaRa} R Rajaraman and R Jackiw {\em Phys. Rev. Lett.} {\bf 54} 
(1985) 1219.
\bibitem{Kim} Y W Kim et al. {\em Z. Phys.} {\bf C 69} (1995) 175.
\bibitem{Asv} A S Vytheeswaran {\em J. Phys.} {\bf G 19} (1993) 957.
\bibitem{Stu} E C G St\"uckelberg {\em Helv. Phys. Act.} {\bf 30} (1957) 
209.
\bibitem{NbRb} N Banerjee and R Banerjee {\em Mod Phys Lett} {\bf A11} 
(1996) 1919; 
Y W Kim, M I Park, Y J Park and S J Yoon, {\em Int J Mod Phys} {\bf A 
12} (1997) 4217, hep-th/9512110.
\bibitem{AsvP} A S Vytheeswaran {\em Int J Mod Phys} {\bf A 13} (1998) 
765, hep-th/9701050.
\bibitem{RaBa} R Banerjee {\em Phys. Rev.} {\bf D 48} (1993) R5467.
\end{thebibliography}
\end{document}